\documentstyle[12pt,fullpage,psfig]{article}

\newenvironment{references}
	{\section{References}\begin{list}{}{
		\topsep=0in
		\partopsep=0in
		\itemsep=0in
		\parsep=0in
		\rightmargin=0in
		\leftmargin=3ex
		\itemindent=-1.0\leftmargin
		\labelsep=0in
		\labelwidth=0in}}
	{\end{list}}

\begin{document}

\begin{center}
{\LARGE\bf Distance measures in cosmology} \\[1.0\baselineskip]
{\large\sc David W. Hogg} \\[0.5\baselineskip]
{\normalsize\sl Institute for Advanced Study, 1 Einstein Drive,
   Princeton NJ 08540} \\
{\normalsize\tt hogg@ias.edu} \\[0.5\baselineskip]
{2000 December}
\end{center}

\section{Introduction}

In cosmology (or to be more specific, {\em cosmography,\/} the
measurement of the Universe) there are many ways to specify the
distance between two points, because in the expanding Universe, the
distances between comoving objects are constantly changing, and
Earth-bound observers look back in time as they look out in distance.
The unifying aspect is that all distance measures somehow measure the
separation between events on radial null trajectories, ie,
trajectories of photons which terminate at the observer.

In this note, formulae for many different cosmological distance
measures are provided.  I treat the concept of ``distance measure''
very liberally, so, for instance, the lookback time and comoving
volume are both considered distance measures.  The bibliography of
source material can be consulted for many of the derivations; this is
merely a ``cheat sheet.''  Minimal $C$ routines (KR) which compute all
of these distance measures are available from the author upon request.
Comments and corrections are highly appreciated, as are
acknowledgments or citation in research that makes use of this summary
or the associated code.

\section{Cosmographic parameters}

The {\em Hubble constant\/} $H_0$ is the constant of proportionality
between recession speed $v$ and distance $d$ in the expanding
Universe;
\begin{equation}
v=H_0\, d
\end{equation}
The subscripted ``0'' refers to the present epoch because in general
$H$ changes with time.  The dimensions of $H_0$ are inverse time, but
it is usually written
\begin{equation}
H_0=100\,h~{\rm km\,s^{-1}\,Mpc^{-1}}
\end{equation}
where $h$ is a dimensionless number parameterizing our ignorance.
(Word on the street is that $0.6<h<0.9$.)  The inverse of the Hubble
constant is the {\em Hubble time\/} $t_{\rm H}$
\begin{equation}
\label{eq:th}
t_{\rm H}\equiv\frac{1}{H_0}
= 9.78\times10^9\,h^{-1}~{\rm yr}= 3.09\times10^{17}\,h^{-1}~{\rm s}
\end{equation}
and the speed of light $c$ times the Hubble time is the {\em Hubble
distance\/} $D_{\rm H}$
\begin{equation}
\label{eq:dh}
D_{\rm H}\equiv\frac{c}{H_0}
= 3000\,h^{-1}~{\rm Mpc}= 9.26\times10^{25}\,h^{-1}~{\rm m}
\end{equation}
These quantities set the scale of the Universe, and often cosmologists
work in geometric units with $c=t_{\rm H}=D_{\rm H}=1$.

The mass density $\rho$ of the Universe and the value of the
cosmological constant $\Lambda$ are dynamical properties of the
Universe, affecting the time evolution of the metric, but in these
notes we will treat them as purely kinematic parameters.  They can be
made into dimensionless density parameters $\Omega_{\rm M}$ and
$\Omega_{\Lambda}$ by
\begin{equation}
\Omega_{\rm M}\equiv\frac{8\pi\,G\,\rho_0}{3\,H_0^2}
\end{equation}
\begin{equation}
\Omega_{\Lambda}\equiv\frac{\Lambda\,c^2}{3\,H_0^2}
\end{equation}
(Peebles, 1993, pp~310--313), where the subscripted ``0''s indicate
that the quantities (which in general evolve with time) are to be
evaluated at the present epoch.  A third density parameter $\Omega_k$
measures the ``curvature of space'' and can be defined by the relation
\begin{equation}
\Omega_{\rm M}+\Omega_{\Lambda}+\Omega_k= 1
\end{equation}
These parameters completely determine the geometry of the Universe if
it is homogeneous, isotropic, and matter-dominated.  By the way, the
critical density $\Omega=1$ corresponds to $7.5\times
10^{21}\,h^{-1}\,M_{\odot}\,D_{\rm H}^{-3}$, where $M_{\odot}$ is the
mass of the Sun.

Most believe that it is in some sense ``unlikely'' that all three of
these density parameters be of the same order, and we know that
$\Omega_{\rm M}$ is significantly larger than zero, so many guess that
$(\Omega_{\rm M},\Omega_{\Lambda},\Omega_k)=(1,0,0)$, with
$(\Omega_{\rm M},1-\Omega_{\rm M},0)$ and $(\Omega_{\rm
M},0,1-\Omega_{\rm M})$ tied for second place.\footnote{This sentence,
unmodified from the first incarnation of these notes, can be used by
historians of cosmology to determine, at least roughly, when they were
written.}  If $\Omega_{\Lambda}=0$, then the {\em deceleration
parameter\/} $q_0$ is just half $\Omega_{\rm M}$, otherwise $q_0$ is
not such a useful parameter.  When I perform cosmographic calculations
and I want to cover all the bases, I use the three world models
\begin{center}
\begin{tabular}{lcc}
name & $\Omega_{\rm M}$ & $\Omega_{\Lambda}$ \\ \hline
Einstein--de-Sitter & 1 & 0 \\
low density & 0.05 & 0 \\
high lambda & 0.2 & 0.8 \\
\end{tabular}
\end{center}
These three models push the observational limits in different
directions.  Some would say that all three of these models are already
ruled out, the first by mass accounting, the second by anisotropies
measured in the cosmic microwave background, and the third by lensing
statistics.  It is fairly likely that the true world model is
somewhere in-between these (unless the $\Omega_{\rm
M},\Omega_{\Lambda},\Omega_k$ parameterization is itself wrong).

\section{Redshift}

The {\em redshift\/} $z$ of an object is the fractional doppler shift
of its emitted light resulting from radial motion
\begin{equation}
z\equiv \frac{\nu_{\rm e}}{\nu_{\rm o}}-1 = \frac{\lambda_{\rm o}}{\lambda_{\rm e}}-1
\end{equation}
where $\nu_{\rm o}$ and $\lambda_{\rm o}$ are the observed frequency
and wavelength, and $\nu_{\rm e}$ and $\lambda_{\rm e}$ are the
emitted.  In special relativity, redshift is related to radial
velocity $v$ by
\begin{equation}
\label{eq:doppler}
1+z = \sqrt{\frac{1+v/c}{1-v/c}}
\end{equation}
where $c$ is the speed of light.  In general relativity,
(\ref{eq:doppler}) is true in one particular coordinate system, but
not any of the traditionally used coordinate systems.  Many feel
(partly for this reason) that it is wrong to view relativistic
redshifts as being due to radial velocities at all (eg, Harrison,
1993).  I do not agree.  On the other hand, redshift is directly
observable and radial velocity is not; these notes concentrate on
observables.

The difference between an object's measured redshift $z_{\rm obs}$ and
its {\em cosmological redshift\/} $z_{\rm cos}$ is due to its (radial)
{\em peculiar velocity\/} $v_{\rm pec}$; ie, we define the
cosmological redshift as that part of the redshift due solely to the
expansion of the Universe, or {\em Hubble flow.\/} The peculiar
velocity is related to the redshift difference by
\begin{equation}
v_{\rm pec} = c\,\frac{(z_{\rm obs}-z_{\rm cos})}{(1+z)}
\end{equation}
where I have assumed $v_{\rm pec}\ll c$.  This can be derived from
(\ref{eq:doppler}) by taking the derivative and using the special
relativity formula for addition of velocities.  From here on, we
assume $z=z_{\rm cos}$.

For small $v/c$, or small distance $d$, in the expanding Universe, the
velocity is linearly proportional to the distance (and all the
distance measures, eg, angular diameter distance, luminosity distance,
etc, converge)
\begin{equation}
z \approx \frac{v}{c} = \frac{d}{D_{\rm H}}
\end{equation}
where $D_{\rm H}$ is the Hubble distance defined in (\ref{eq:dh}).
But this is {\em only true for small redshifts!\/} It is important to
note that many galaxy redshift surveys, when presenting redshifts as
radial velocities, {\em always\/} use the non-relativistic
approximation $v=c\,z$, even when it may not be physically appropriate
(eg, Fairall 1992).

In terms of cosmography, the cosmological redshift is directly related
to the scale factor $a(t)$, or the ``size'' of the Universe.  For an
object at redshift $z$
\begin{equation}
1+z = \frac{a(t_{\rm o})}{a(t_{\rm e})}
\end{equation}
where $a(t_{\rm o})$ is the size of the Universe at the time the light
from the object is observed, and $a(t_{\rm e})$ is the size at the
time it was emitted.

Redshift is almost always determined with respect to us (or the frame
centered on us but stationary with respect to the microwave
background), but it is possible to define the redshift $z_{12}$
between objects 1 and 2, both of which are cosmologically redshifted
relative to us: the redshift $z_{12}$ of an object at redshift $z_2$
relative to a hypothetical observer at redshift $z_1<z_2$ is given by
\begin{equation}
1+z_{12} = \frac{a(t_1)}{a(t_2)} = \frac{1+z_2}{1+z_1}
\end{equation}

\section{Comoving distance (line-of-sight)}

A small {\em comoving distance\/} $\delta D_{\rm C}$ between two
nearby objects in the Universe is the distance between them which
remains constant with epoch if the two objects are moving with the
Hubble flow.  In other words, it is the distance between them which
would be measured with rulers at the time they are being observed (the
{\em proper distance}) divided by the ratio of the scale factor of the
Universe then to now; it is the proper distance multiplied by $(1+z)$.
The total line-of-sight comoving distance $D_{\rm C}$ from us to a
distant object is computed by integrating the infinitesimal $\delta
D_{\rm C}$ contributions between nearby events along the radial ray
from $z=0$ to the object.

Following Peebles (1993, pp~310--321) (who calls the transverse
comoving distance by the confusing name ``angular size distance,''
which is {\em not\/} the same as ``angular diameter distance''
introduced below), we define the function
\begin{equation}
\label{eq:ez}
E(z)\equiv\sqrt{\Omega_{\rm M}\,(1+z)^3+\Omega_k\,(1+z)^2+\Omega_{\Lambda}}
\end{equation}
which is proportional to the time derivative of the logarithm of the
scale factor (ie, $\dot{a}(t)/a(t)$), with $z$ redshift and
$\Omega_{\rm M}$, $\Omega_k$ and $\Omega_{\Lambda}$ the three density
parameters defined above.  (For this reason, $H(z)=H_0\,E(z)$ is the
Hubble constant as measured by a hypothetical astronomer working at
redshift $z$.)  Since $dz=da$, $dz/E(z)$ is proportional to the
time-of-flight of a photon traveling across the redshift interval
$dz$, divided by the scale factor at that time.  Since the speed of
light is constant, this is a proper distance divided by the scale
factor, which is the definition of a comoving distance.  The total
line-of-sight comoving distance is then given by integrating these
contributions, or
\begin{equation}
D_{\rm C} = D_{\rm H}\,\int_0^z\frac{dz'}{E(z')}
\end{equation}
where $D_{\rm H}$ is the Hubble distance defined by (\ref{eq:dh}).

In some sense the line-of-sight comoving distance is the fundamental
distance measure in cosmography since, as will be seen below, all
others are quite simply derived in terms of it.  The line-of-sight
comoving distance between two nearby events (ie, close in redshift or
distance) is the distance which we would measure locally between the
events today if those two points were locked into the Hubble flow.  It
is the correct distance measure for measuring aspects of large-scale
structure imprinted on the Hubble flow, eg, distances between
``walls.''

\section{Comoving distance (transverse)}

The comoving distance between two events at the same redshift or
distance but separated on the sky by some angle $\delta\theta$ is
$D_{\rm M}\,\delta\theta$ and the transverse comoving distance $D_{\rm
M}$ (so-denoted for a reason explained below) is simply related to the
line-of-sight comoving distance $D_{\rm C}$:
\begin{equation}
D_{\rm M} = \left\{
\begin{array}{ll}
D_{\rm H}\,\frac{1}{\sqrt{\Omega_k}}\,\sinh\left[\sqrt{\Omega_k}\,D_{\rm C}/D_{\rm H}\right] & {\rm for}~\Omega_k>0 \\
D_{\rm C} & {\rm for}~\Omega_k=0 \\
D_{\rm H}\,\frac{1}{\sqrt{|\Omega_k|}}\,\sin\left[\sqrt{|\Omega_k|}\,D_{\rm C}/D_{\rm H}\right] & {\rm for}~\Omega_k<0
\end{array}
\right.
\end{equation}
where the trigonometric functions $\sinh$ and $\sin$ account for what
is called ``the curvature of space.''  (Space curvature is not
coordinate-free; a change of coordinates makes space flat; the only
coordinate-free curvature is space--time curvature, which is related
to the local mass--energy density or really stress--energy tensor.)
For $\Omega_{\Lambda}=0$, there is an analytic solution to the
equations
\begin{equation}
D_{\rm M}=D_{\rm H}\,\frac{2\,[2-\Omega_{\rm M}\,(1-z)-
(2-\Omega_{\rm M})\,\sqrt{1+\Omega_{\rm M}\,z}]}{\Omega_{\rm M}^2\,(1+z)}
~{\rm for}~\Omega_{\Lambda}=0
\end{equation}
(Weinberg, 1972, p.~485; Peebles, 1993, pp~320--321).  Some (eg,
Weedman, 1986, pp~59--60) call this distance measure ``proper
distance,'' which, though common usage, is bad style.\footnote{The
word ``proper'' has a specific use in relativity.  The {\em proper
time\/} between two nearby events is the time delay between the events
in the frame in which they take place at the same location, and the
{\em proper distance\/} between two nearby events is the distance
between them in the frame in which they happen at the same time.  In
the cosmological context, it is the distance measured by a ruler at
the time of observation.  The transverse comoving distance $D_{\rm M}$
is {\em not\/} a proper distance---it is a proper distance divided by
a ratio of scale factors.}

(Although these notes follow the Peebles derivation, there is a
qualitatively distinct method using what is known as the {\em
development angle\/} $\chi$, which increases as the Universe evolves.
This method is generally preferred by relativists; eg, Misner, Thorne
\& Wheeler 1973, pp~782--785).

\begin{figure}
\psfig{figure=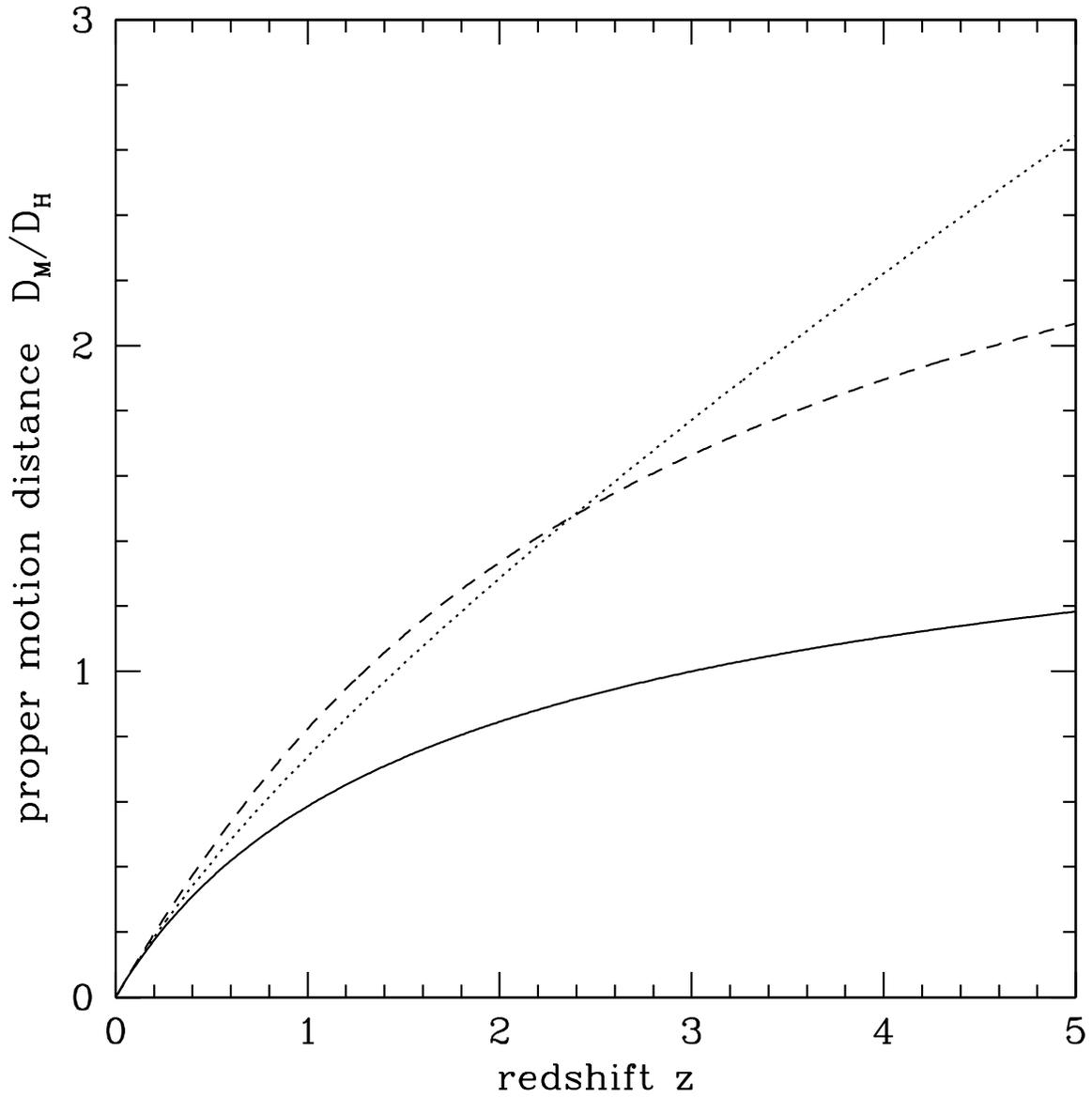,width=\textwidth}
\caption[The dimensionless proper motion distance $D_{\rm M}/D_{\rm
H}$.]{ The dimensionless proper motion distance $D_{\rm M}/D_{\rm H}$.
The three curves are for the three world models, Einstein-de~Sitter
$(\Omega_{\rm M},\Omega_{\Lambda})=(1,0)$, solid; low-density,
$(0.05,0)$, dotted; and high lambda, $(0.2,0.8)$, dashed.}
\label{propmotdis}
\end{figure}

The comoving distance happens to be equivalent to the {\em proper
motion distance\/} (hence the name $D_{\rm M}$), defined as the ratio
of the actual transverse velocity (in distance over time) of an object
to its proper motion (in radians per unit time) (Weinberg, 1972,
pp~423--424).  The proper motion distance is plotted in
Figure~\ref{propmotdis}.  Proper motion distance is used, for example,
in computing radio jet velocities from knot motion.

\section{Angular diameter distance}

\begin{figure}
\psfig{figure=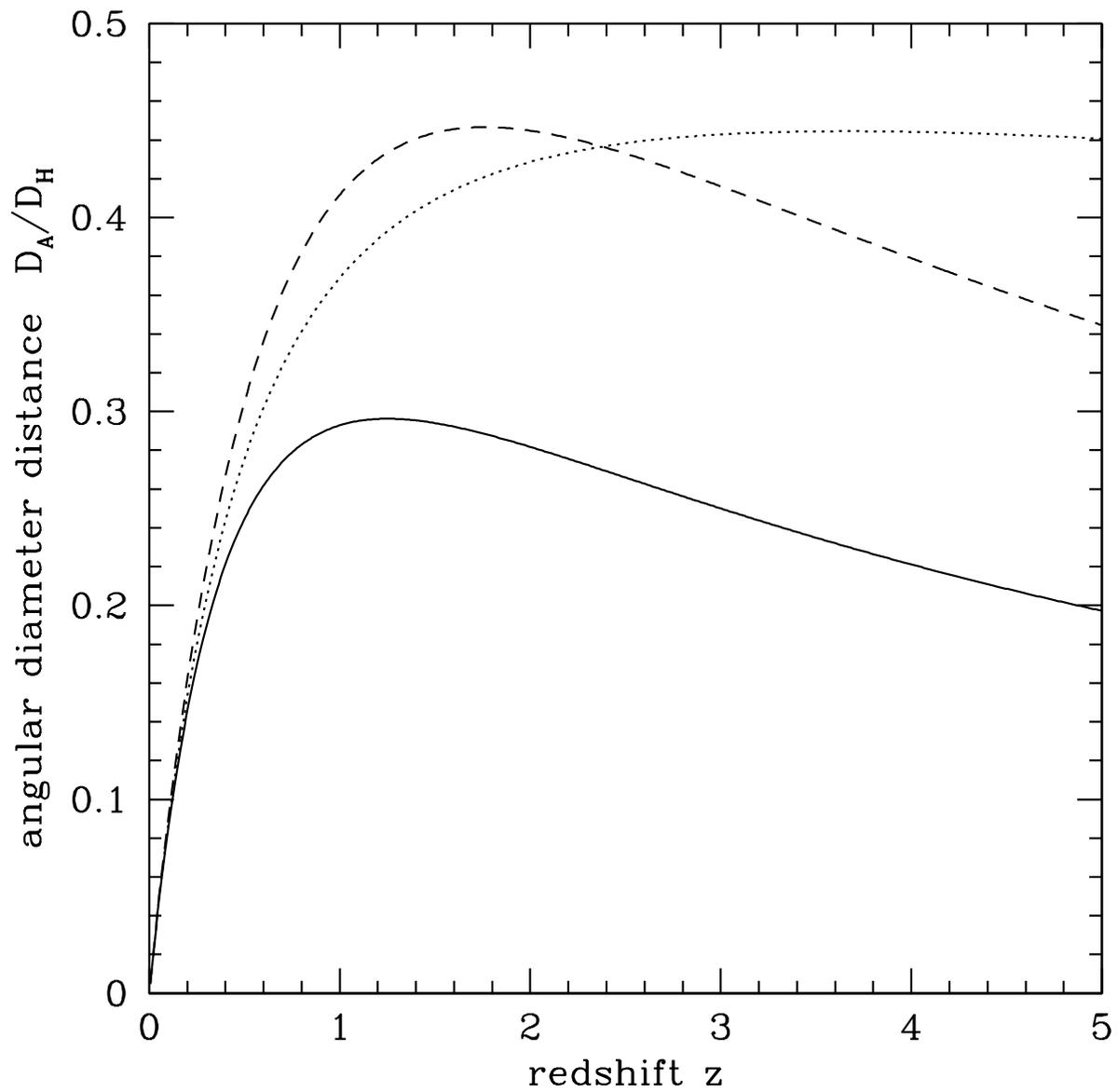,width=\textwidth}
\caption[The dimensionless angular diameter distance $D_{\rm A}/D_{\rm
H}$.]{ The dimensionless angular diameter distance $D_{\rm A}/D_{\rm
H}$.  The three curves are for the three world models, $(\Omega_{\rm
M},\Omega_{\Lambda})=(1,0)$, solid; $(0.05,0)$, dotted; and
$(0.2,0.8)$, dashed.}
\label{angdidis}
\end{figure}

The {\em angular diameter distance\/} $D_{\rm A}$ is defined as the
ratio of an object's physical transverse size to its angular size (in
radians).  It is used to convert angular separations in telescope
images into proper separations at the source.  It is famous for not
increasing indefinitely as $z\rightarrow\infty$; it turns over at
$z\sim 1$ and thereafter more distant objects actually appear larger
in angular size.  Angular diameter distance is related to the
transverse comoving distance by
\begin{equation}
D_{\rm A} = \frac{D_{\rm M}}{1+z}
\end{equation}
(Weinberg, 1972, pp~421--424; Weedman, 1986, pp~65--67; Peebles, 1993,
pp~325--327).  The angular diameter distance is plotted in
Figure~\ref{angdidis}.  At high redshift, the angular diameter
distance is such that 1~arcsec is on the order of 5~kpc.

There is also an angular diameter distance $D_{\rm A12}$ between two
objects at redshifts $z_1$ and $z_2$, frequently used in gravitational
lensing.  It is {\em not\/} found by subtracting the two individual
angular diameter distances!  The correct formula, for $\Omega_k\geq
0$, is
\begin{equation}
D_{\rm A12}= \frac{1}{1+z_2}\,\left[
 D_{\rm M2}\,\sqrt{1+\Omega_k\,\frac{D_{\rm M1}^2}{D_{\rm H}^2}}
 - D_{\rm M1}\,\sqrt{1+\Omega_k\,\frac{D_{\rm M2}^2}{D_{\rm H}^2}}\right]
\end{equation}
where $D_{\rm M1}$ and $D_{\rm M2}$ are the transverse comoving
distances to $z_1$ and $z_2$, $D_{\rm H}$ is the Hubble distance, and
$\Omega_k$ is the curvature density parameter (Peebles, 1993,
pp~336--337).  Unfortunately, the above formula is {\em not correct\/}
for $\Omega_k<0$ (Phillip Helbig, 1998, private communication).

\section{Luminosity distance}

\begin{figure}
\psfig{figure=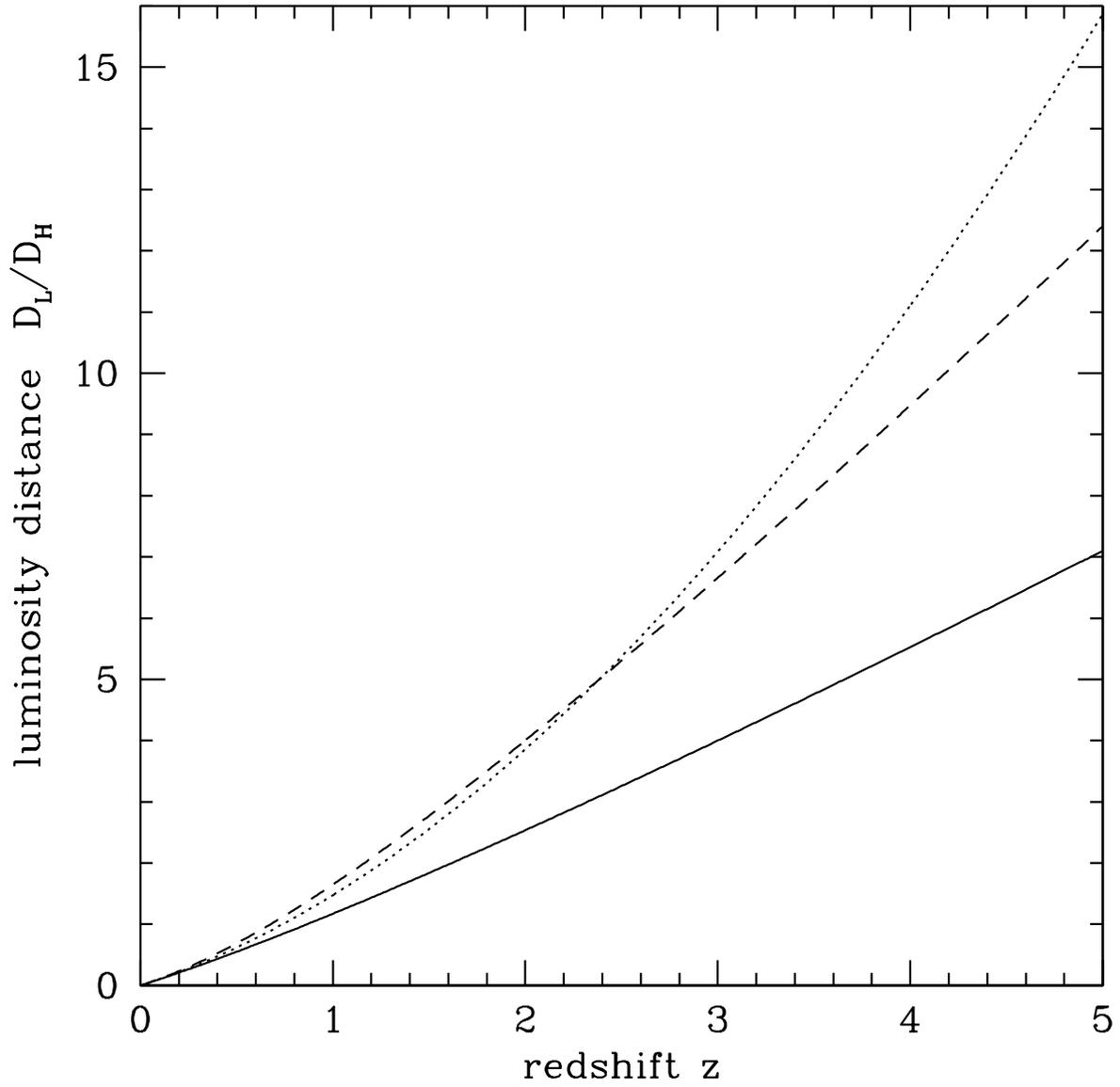,width=\textwidth}
\caption[The dimensionless luminosity distance $D_{\rm L}/D_{\rm
H}$.]{ The dimensionless luminosity distance $D_{\rm L}/D_{\rm H}$.
The three curves are for the three world models, $(\Omega_{\rm
M},\Omega_{\Lambda})=(1,0)$, solid; $(0.05,0)$, dotted; and
$(0.2,0.8)$, dashed.}
\label{lumdis}
\end{figure}

The {\em luminosity distance\/} $D_{\rm L}$ is defined by the
relationship between bolometric (ie, integrated over all frequencies)
flux $S$ and bolometric luminosity $L$:
\begin{equation}
D_{\rm L}\equiv \sqrt{\frac{L}{4\pi\,S}}
\end{equation}
It turns out that this is related to the transverse comoving distance
and angular diameter distance by
\begin{equation}
D_{\rm L} = (1+z)\,D_{\rm M} = (1+z)^2\,D_{\rm A}
\end{equation}
(Weinberg, 1972, pp~420--424; Weedman, 1986, pp~60--62).  The latter
relation follows from the fact that the surface brightness of a
receding object is reduced by a factor $(1+z)^{-4}$, and the angular
area goes down as $D_{\rm A}^{-2}$.  The luminosity distance is
plotted in Figure~\ref{lumdis}.

If the concern is not with bolometric quantities but rather with
differential flux $S_{\nu}$ and luminosity $L_{\nu}$, as is usually
the case in astronomy, then a correction, the {\em k-correction,\/}
must be applied to the flux or luminosity because the redshifted
object is emitting flux in a different band than that in which you are
observing.  The k-correction depends on the spectrum of the object in
question, and is unnecessary only if the object has spectrum
$\nu\,L_{\nu}={\rm constant}$.  For any other spectrum the
differential flux $S_{\nu}$ is related to the differential luminosity
$L_{\nu}$ by
\begin{equation}
S_{\nu} = (1+z)\,\frac{L_{(1+z)\nu}}{L_{\nu}}\,\frac{L_{\nu}}{4\pi\,D_{\rm L}^2}
\end{equation}
where $z$ is the redshift, the ratio of luminosities equalizes the
difference in flux between the observed and emitted bands, and the
factor of $(1+z)$ accounts for the redshifting of the bandwidth.
Similarly, for differential flux per unit wavelength,
\begin{equation}
S_{\lambda} = \frac{1}{(1+z)}\,\frac{L_{\lambda/(1+z)}}{L_{\lambda}}\,
\frac{L_{\lambda}}{4\pi\,D_{\rm L}^2}
\end{equation}
(Peebles, 1993, pp~330--331; Weedman, 1986, pp~60--62).  In this
author's opinion, the most natural flux unit is differential flux per
unit log frequency or log wavelength
$\nu\,S_{\nu}=\lambda\,S_{\lambda}$ for which there is no redshifting
of the bandpass so
\begin{equation}
\nu\,S_{\nu} = \frac{\nu_{\rm e}\,L_{\nu_{\rm e}}}{4\pi\,D_{\rm L}^2}
\end{equation}
where $\nu_{\rm e}=(1+z)\nu$ is the emitted frequency.  These
equations are straightforward to generalize to bandpasses of finite
width.

\begin{figure}
\psfig{figure=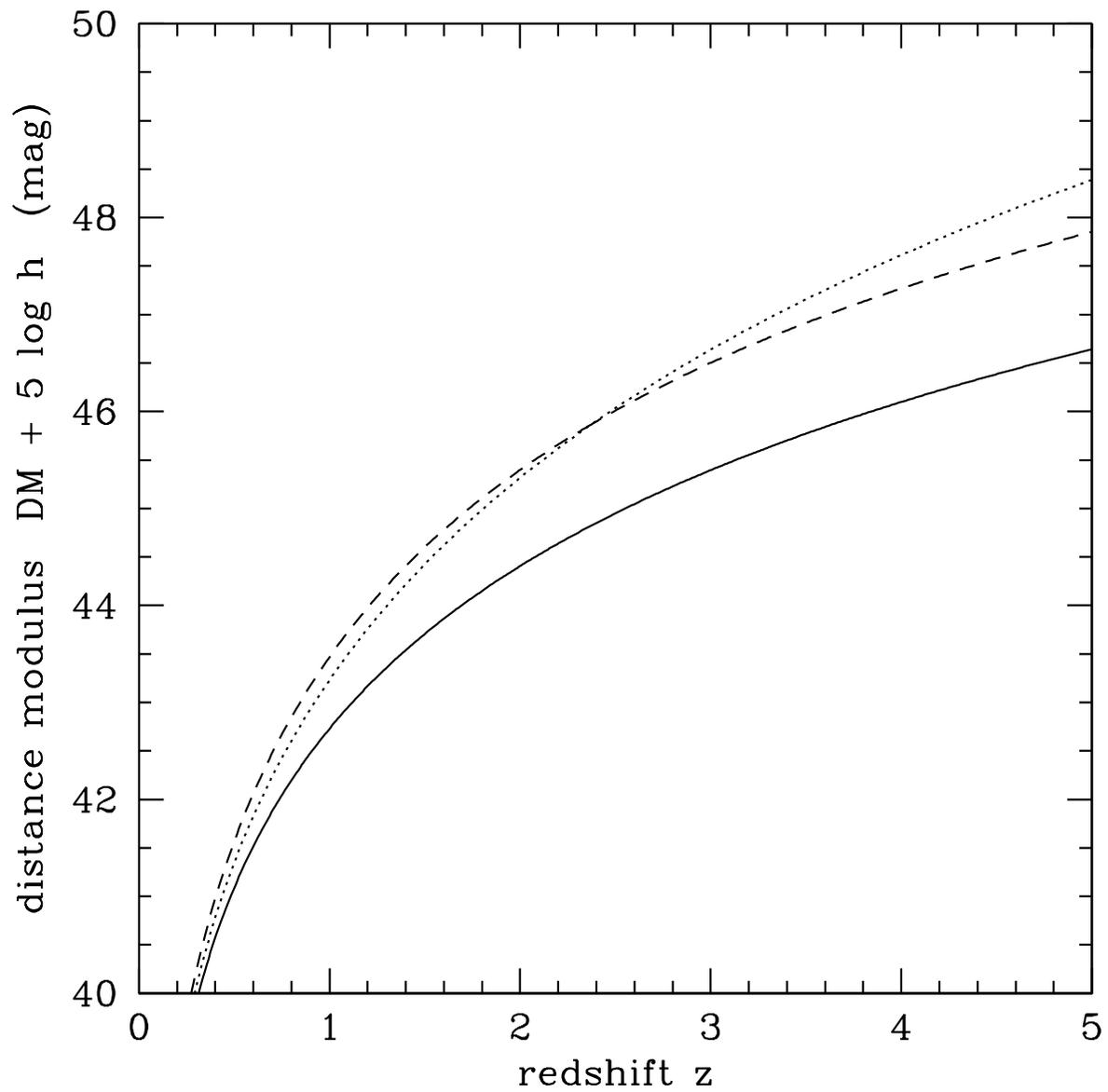,width=\textwidth}
\caption[The distance modulus $DM$.]{ The distance modulus $DM$.  The
three curves are for the three world models, $(\Omega_{\rm
M},\Omega_{\Lambda})=(1,0)$, solid; $(0.05,0)$, dotted; and
$(0.2,0.8)$, dashed.}
\label{distmod}
\end{figure}

The {\em apparent magnitude\/} $m$ of an astronomical source in a
photometric bandpass is defined to be the ratio of the apparent flux
of that source to the apparent flux of the bright star Vega, through
that bandpass ({\em don't\/} ask me about ``AB magnitudes'').  The
{\em distance modulus\/} $DM$ is defined by
\begin{equation}
DM\equiv 5\,\log \left(\frac{D_{\rm L}}{10~{\rm pc}}\right)
\end{equation}
because it is the magnitude difference between an object's observed
bolometric flux and what it would be if it were at $10~{\rm pc}$ (this
was once thought to be the distance to Vega).  The distance modulus is
plotted in Figure~\ref{distmod}.  The absolute magnitude $M$ is the
astronomer's measure of luminosity, defined to be the apparent
magnitude the object in question would have if it were at 10~pc, so
\begin{equation}
m=M+DM+K
\end{equation}
where $K$ is the k-correction
\begin{equation}
K = -2.5\,\log\left[(1+z)\,\frac{L_{(1+z)\nu}}{L_{\nu}}\right]=
-2.5\,\log \left[\frac{1}{(1+z)}\,\frac{L_{\lambda/(1+z)}}{L_{\lambda}}\right]
\end{equation}
(eg, Oke \& Sandage, 1968).

\section{Parallax distance}

If it were possible to measure parallaxes for high redshift objects,
the distance so measured would be the {\em parallax distance\/}
$D_{\rm P}$ (Weinberg, 1972, pp~418--420).  It may be possible, one
day, to measure parallaxes to distant galaxies using gravitational
lensing, although in these cases, a modified parallax distance is used
which takes into account the redshifts of both the source and the lens
(Schneider, Ehlers \& Falco, 1992, pp~508--509), a discussion of which
is beyond the scope of these notes.

\section{Comoving volume}

\begin{figure}
\psfig{figure=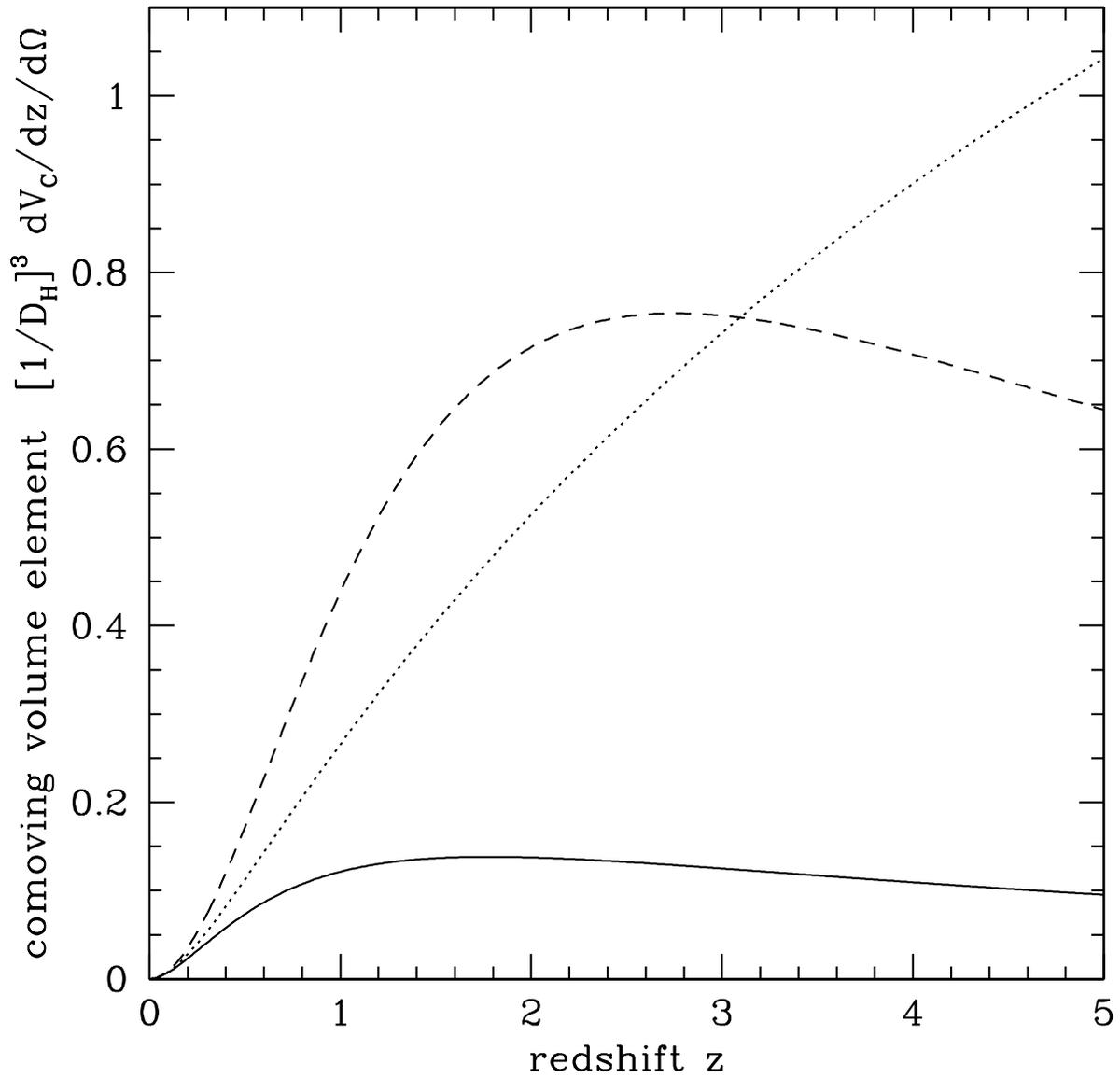,width=\textwidth}
\caption[The dimensionless comoving volume element $(1/D_{\rm
H})^3\,(dV_{\rm C}/dz)$.]{ The dimensionless comoving volume element
$(1/D_{\rm H})^3\,(dV_{\rm C}/dz)$.  The three curves are for the
three world models, $(\Omega_{\rm M},\Omega_{\Lambda})=(1,0)$, solid;
$(0.05,0)$, dotted; and $(0.2,0.8)$, dashed.}
\label{dcomvoldz}
\end{figure}

The {\em comoving volume\/} $V_{\rm C}$ is the volume measure in which
number densities of non-evolving objects locked into Hubble flow are
constant with redshift.  It is the proper volume times three factors
of the relative scale factor now to then, or $(1+z)^3$.  Since the
derivative of comoving distance with redshift is $1/E(z)$ defined in
(\ref{eq:ez}), the angular diameter distance converts a solid angle
$d\Omega$ into a proper area, and two factors of $(1+z)$ convert a
proper area into a comoving area, the comoving volume element in solid
angle $d\Omega$ and redshift interval $dz$ is
\begin{equation}
dV_{\rm C}= D_{\rm H}\,\frac{(1+z)^2\,D_{\rm A}^2}{E(z)}\,d\Omega\,dz
\end{equation}
where $D_{\rm A}$ is the angular diameter distance at redshift $z$ and
$E(z)$ is defined in (\ref{eq:ez}) (Weinberg, 1972, p.~486; Peebles,
1993, pp~331--333).  The comoving volume element is plotted in
Figure~\ref{dcomvoldz}.  The integral of the comoving volume element
from the present to redshift $z$ gives the total comoving volume,
all-sky, out to redshift $z$
\begin{equation}
V_{\rm C} = \left\{
\begin{array}{ll}
  \left(\frac{4\pi\,D_{\rm H}^3}{2\,\Omega_k}\right)\,
  \left[\frac{D_{\rm M}}{D_{\rm H}}\,
  \sqrt{1+\Omega_k\,\frac{D_{\rm M}^2}{D_{\rm H}^2}}
  -\frac{1}{\sqrt{|\Omega_k|}}\,
  {\rm arcsinh}\left(\sqrt{|\Omega_k|}\,\frac{D_{\rm M}}{D_{\rm H}}\right)\right]
  & {\rm for}~\Omega_k>0 \\
  \frac{4\pi}{3}\,D_{\rm M}^3
  & {\rm for}~\Omega_k=0 \\
  \left(\frac{4\pi\,D_{\rm H}^3}{2\,\Omega_k}\right)\,
  \left[\frac{D_{\rm M}}{D_{\rm H}}\,
  \sqrt{1+\Omega_k\,\frac{D_{\rm M}^2}{D_{\rm H}^2}}
  -\frac{1}{\sqrt{|\Omega_k|}}\,
  {\rm arcsin}\left(\sqrt{|\Omega_k|}\,\frac{D_{\rm M}}{D_{\rm H}}\right)\right]
  & {\rm for}~\Omega_k<0
\end{array}
\right.
\end{equation}
(Carrol, Press \& Turner, 1992), where $D_{\rm H}^3$ is sometimes
called the {\em Hubble volume.\/} The comoving volume element and its
integral are both used frequently in predicting number counts or
luminosity densities.

\section{Lookback time}

\begin{figure}
\psfig{figure=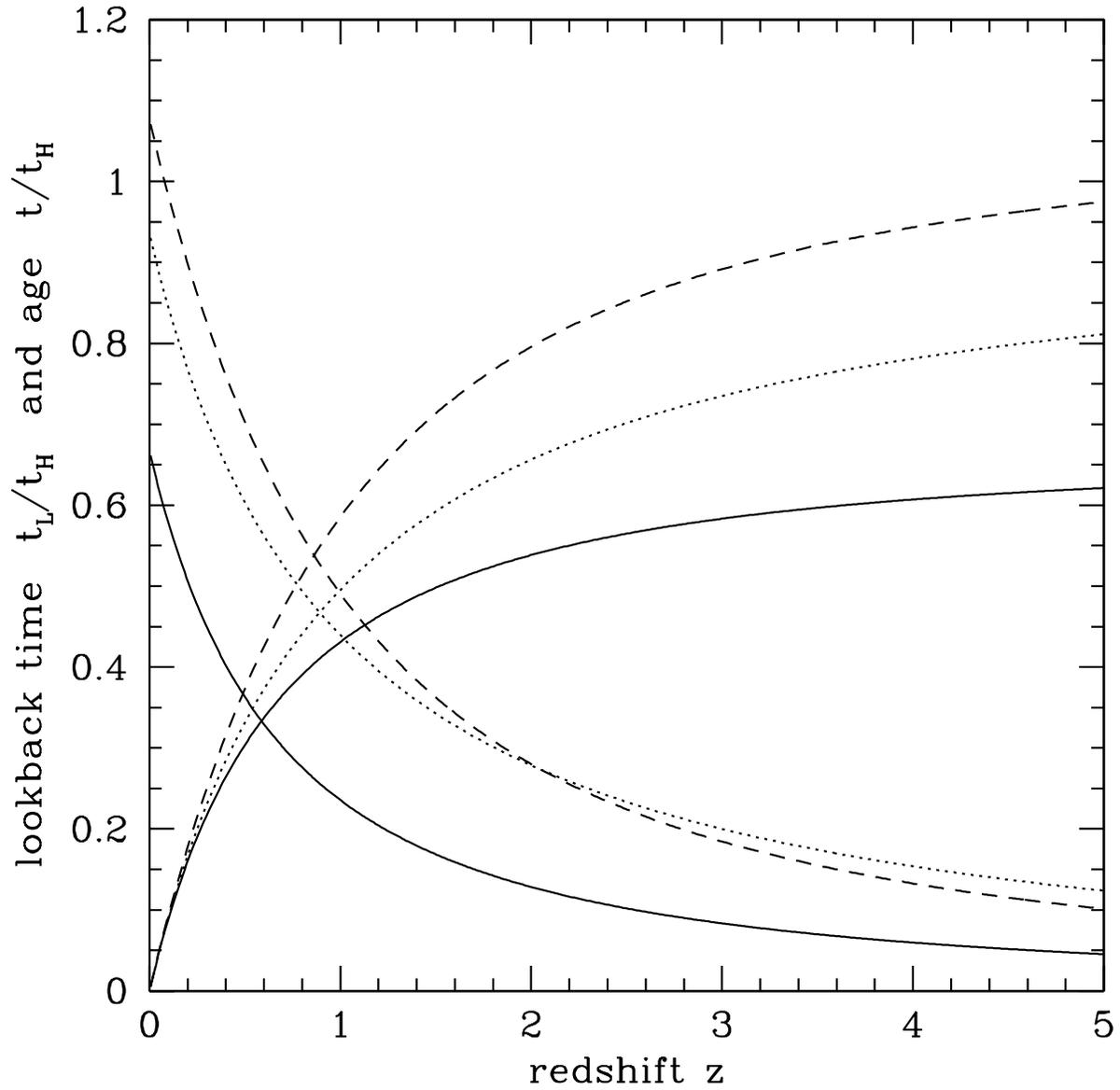,width=\textwidth}
\caption[The dimensionless lookback time $t_{\rm L}/t_{\rm H}$ and age
$t/t_{\rm H}$.]{The dimensionless lookback time $t_{\rm L}/t_{\rm H}$
and age $t/t_{\rm H}$.  Curves cross at the redshift at which the
Universe is half its present age.  The three curves are for the three
world models, $(\Omega_{\rm M},\Omega_{\Lambda})=(1,0)$, solid;
$(0.05,0)$, dotted; and $(0.2,0.8)$, dashed.}
\label{lookback}
\end{figure}

The {\em lookback time\/} $t_{\rm L}$ to an object is the difference
between the age $t_{\rm o}$ of the Universe now (at observation) and
the age $t_{\rm e}$ of the Universe at the time the photons were
emitted (according to the object).  It is used to predict properties
of high-redshift objects with evolutionary models, such as passive
stellar evolution for galaxies.  Recall that $E(z)$ is the time
derivative of the logarithm of the scale factor $a(t)$; the scale
factor is proportional to $(1+z)$, so the product $(1+z)\,E(z)$ is
proportional to the derivative of $z$ with respect to the lookback
time, or
\begin{equation}
t_{\rm L} = t_{\rm H}\,\int_0^z \frac{dz'}{(1+z')\,E(z')}
\end{equation}
(Peebles, 1993, pp~313--315; Kolb \& Turner 1990, pp~52--56, give some
analytic solutions to this equation, but they are concerned with the
age $t(z)$, so they integrate from $z$ to $\infty$).  The lookback
time and age are plotted in Figure~\ref{lookback}.

\section{Probability of intersecting objects}
\label{sec:optdepth}

\begin{figure}
\psfig{figure=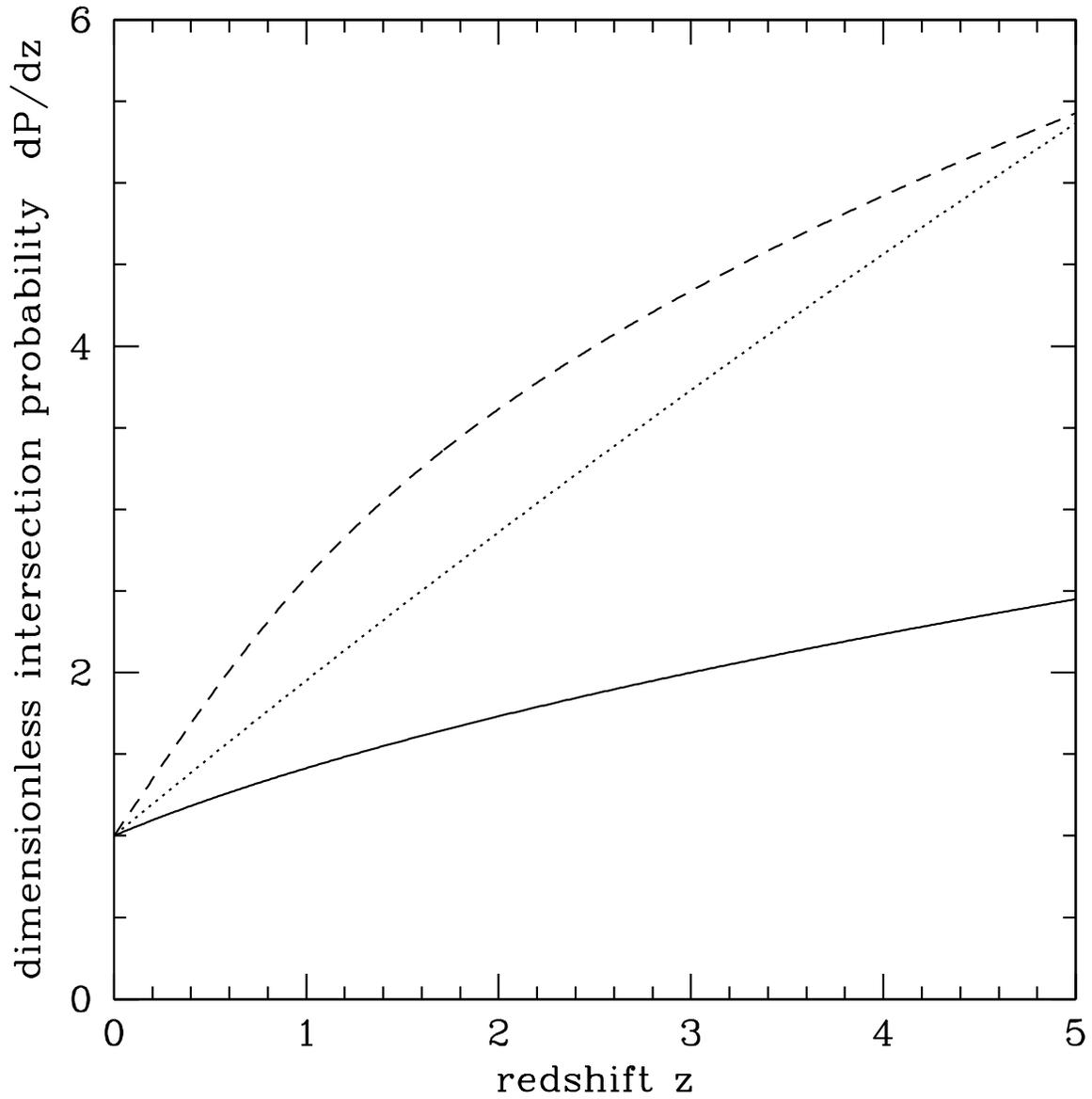,width=\textwidth}
\caption[The dimensionless differential intersection probability
$dP/dz$.]{ The dimensionless differential intersection probability
$dP/dz$; dimensionless in the sense of $n(z)\,\sigma(z)\,D_{\rm H}=1$.
The three curves are for the three world models, $(\Omega_{\rm
M},\Omega_{\Lambda})=(1,0)$, solid; $(0.05,0)$, dotted; and
$(0.2,0.8)$, dashed.}
\label{doptdepthdz}
\end{figure}

Given a population of objects with comoving number density $n(z)$
(number per unit volume) and cross section $\sigma(z)$ (area), what is
the incremental probability $dP$ that a line of sight will intersect
one of the objects in redshift interval $dz$ at redshift $z$?
Questions of this form are asked frequently in the study of QSO
absorption lines or pencil-beam redshift surveys.  The answer is
\begin{equation}
dP=n(z)\,\sigma(z)\,D_{\rm H}\,\frac{(1+z)^2}{E(z)}\,dz
\end{equation}
(Peebles, 1993, pp~323--325).  The dimensionless differential
intersection probability is plotted in Figure~\ref{doptdepthdz}.

\section*{Acknowledgments}
\addcontentsline{toc}{section}{Acknowledgments}

Roger Blandford, Ed Farhi, Jim Peebles and Wal Sargent all contributed
generously to my understanding of this material and Kurt Adelberger,
Lee Armus, Andrew Baker, Deepto Chakrabarty, Alex Filippenko, Andrew
Hamilton, Phillip Helbig, Wayne Hu, John Huchra, Daniel Mortlock, Tom
Murphy, Gerry Neugebauer, Adam Riess, Paul Schechter, Douglas Scott
and Ned Wright caught errors, suggested additional material, or helped
me with wording, conventions or terminology.  I thank the NSF and NASA
for financial support.

\clearpage

\end{document}